%% file: UIC.tex
\documentclass[sigconf]{acmart}
\usepackage{amsmath,amssymb,amsfonts,bm,amsthm}
\usepackage[linesnumbered,ruled]{algorithm2e}
\usepackage{graphicx}
\usepackage{epsfig}
\usepackage[tight,normal]{subfigure}
\usepackage{multirow,makecell,url,footmisc}
\usepackage{enumerate}

\usepackage{textcomp}
\usepackage{booktabs} % For formal tables
\usepackage{subfigure}
\usepackage{amssymb}
\usepackage{amsmath}
\usepackage{booktabs}
\usepackage{bm}
\usepackage{mathtools}
\usepackage{threeparttable}

\usepackage[american]{babel}

\usepackage{color}
\usepackage{threeparttable}
\usepackage{appendix}
\usepackage{comment}
\usepackage{multirow}

\usepackage{blindtext}
\usepackage{comment}
\usepackage{balance}

\usepackage{comment}

\def\BibTeX{{\rm B\kern-.05em{\sc i\kern-.025em b}\kern-.08emT\kern-.1667em\lower.7ex\hbox{E}\kern-.125emX}}

\begin{document}

\copyrightyear{2019} 
\acmYear{2019} 
\setcopyright{acmcopyright}
\acmConference[KDD '19]{The 25th ACM SIGKDD Conference on Knowledge Discovery and Data Mining}{August 4--8, 2019}{Anchorage, AK, USA}
\acmBooktitle{The 25th ACM SIGKDD Conference on Knowledge Discovery and Data Mining (KDD '19), August 4--8, 2019, Anchorage, AK, USA}
\acmPrice{15.00}
\acmDOI{10.1145/3292500.3330666}
\acmISBN{978-1-4503-6201-6/19/08}

%\fancyhead{}
\pagestyle{fancy}
%\rhead｛\thepage｝
\cfoot{\thepage}
\settopmatter{printacmref=true}

\title{Practice on Long Sequential User Behavior Modeling for Click-Through Rate Prediction}

\author{Qi Pi, Weijie Bian, Guorui Zhou, Xiaoqiang Zhu, Kun Gai}
\authornote{Q. Pi and W. Bian share the co-first authorship. Corresponding author is G. Zhou.} 
\affiliation{%
  \institution{Alibaba Group}
   \city{Beijing}
   \country{P.R.China}
}
\email{{piqi.pq, weijie.bwj, guorui.xgr, xiaoqiang.zxq, jingshi.gk}@alibaba-inc.com}

\renewcommand{\shortauthors}{Pi and Bian, et al.}

\begin{abstract}

Click-through rate (CTR) prediction is critical for industrial applications such as recommender system and online advertising. Practically, it plays an important role for CTR modeling in these applications by mining user interest from rich historical behavior data. Driven by the development of deep learning, deep CTR models with ingeniously designed architecture for user interest modeling have been proposed, bringing remarkable improvement of model performance over offline metric.   
However, great efforts are needed to deploy these complex models to online serving system for realtime inference, facing massive traffic request. Things turn to be more difficult when it comes to long sequential user behavior data, as the system latency and storage cost increase approximately linearly with the length of user behavior sequence. 

In this paper, we face directly the challenge of long sequential user behavior modeling and introduce our hands-on practice with the co-design of machine learning algorithm and online serving system for CTR prediction task. 
(i) From serving system view, we decouple the most resource-consuming part of user interest modeling from the entire model by designing a separate module named UIC (User Interest Center). UIC maintains the latest interest state for each user, whose update depends on realtime user behavior trigger event, rather than on traffic request. Hence UIC is latency free for realtime CTR prediction.  
(ii) From machine learning algorithm view, we propose a novel memory-based architecture named MIMN (Multi-channel user Interest Memory Network) to capture user interests from long sequential behavior data, achieving superior performance over state-of-the-art models. MIMN is implemented in an incremental manner with UIC module.    

Theoretically, the co-design solution of UIC and MIMN enables us to handle the user interest modeling with unlimited length of sequential behavior data. Comparison between model performance and system efficiency proves the effectiveness of proposed solution. To our knowledge, this is one of the first industrial solutions that are capable of handling long sequential user behavior data with length scaling up to thousands. It now has been deployed in the display advertising system in Alibaba.

\end{abstract}



\maketitle
\input{ch_intro}

\input{ch_relatedwork}
\input{ch_sysview_pq}

\input{ch_approach_gr}

\input{ch_exp}

\input{ch_concln}

\begin{acks}
The authors would like to thank  for Guoqiang Ma, Zhenzhong Shen, Haoyu Liu,  Weizhao Wang, Chi Ma, Juntao Song, Pengtao Yi who did the really hard work for online system implement.
\end{acks}

\bibliographystyle{ACM-Reference-Format}
\balance
\bibliography{UIC}
\end{document}

%% file: ch_intro.tex
\section{Introduction}

Growing internet takes us to a digital world with personalized online services. Huge amount of user behavior data collected from online systems provides us with great opportunity to better understand user preferences. Technically, capturing user interests from the rich behavior data is critically important since it contributes remarkable improvement for typical real-world applications, such as recommender system and online advertising \cite{zhou2018deep,zhou2019dien,youtube:recommend}. 
In this paper, we limit ourselves to the task of Click-Through Rate (CTR) prediction modeling, which plays a critical role in online services. Solutions discussed here are also applicable in many related tasks, such as conversion rate prediction and user preference modeling.   

Driven by the advance of deep learning, deep CTR models with ingeniously designed architecture for user interest modeling have been proposed, achieving state-of-the-art. 
These models can be roughly divided into two categories: (i) pooling-based architecture \cite{widedeep,youtube:recommend,zhou2018deep} which treats  historical behaviors of users as independent signals and applies sum/max/attention etc. pooling operations to summarize user interest representation, (ii) sequential-modeling architecture \cite{zhou2019dien,quadrana2017gru4rec} which treats user behaviors as sequential signals and applies LSTM/GRU operations for user interest summarization.

\begin{figure}[!htb]
\centering
\setlength{\abovecaptionskip}{0.03cm}
\setlength{\belowcaptionskip}{-0.45cm}
\includegraphics[height=5in, width=3.3in, keepaspectratio]{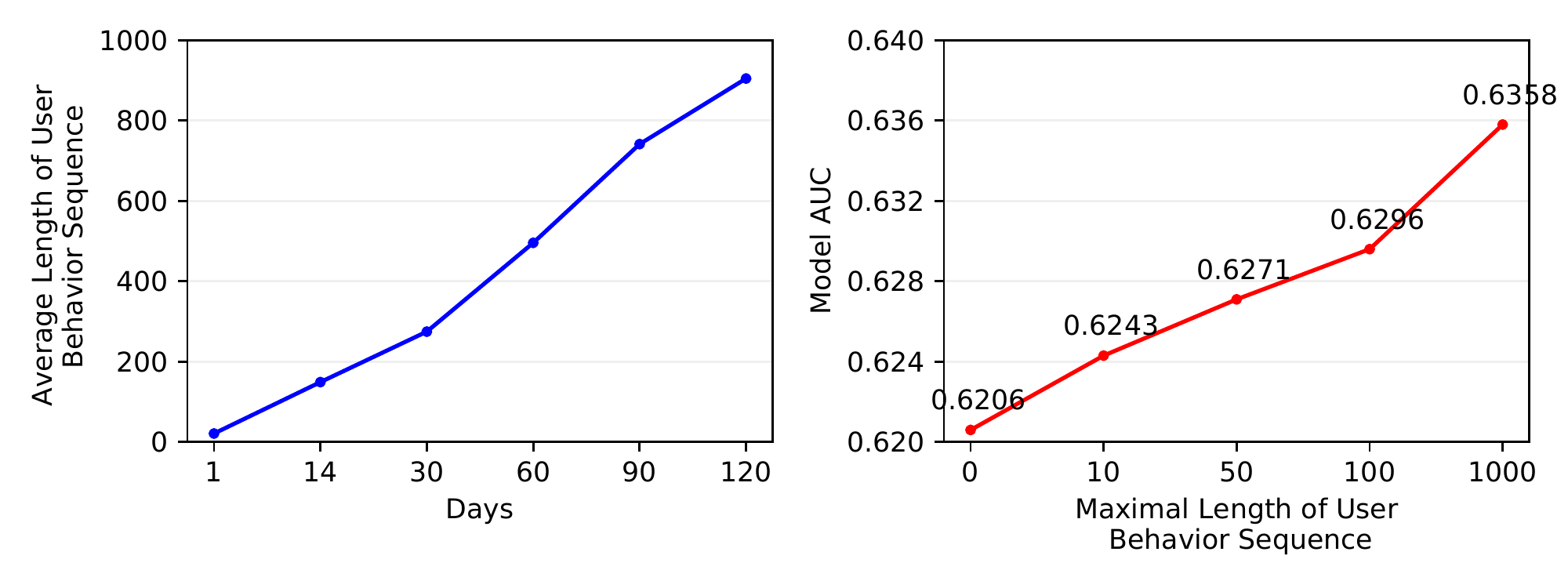}
\caption{Statistics of sequential user behavior data and corresponding model performance in the display advertising system in Alibaba.}
\label{fig:ub}
\end{figure}

However, in industrial applications it needs great effort to deploy these complex models to online serving system for realtime inference, with hundreds of millions of users visiting the system everyday. Things turn to be more difficult when encountering with extremely long sequential user behavior data, because all the aforementioned models need to store the whole user behavior sequences, a.k.a. features, in the online serving system and fetch them to compute the interest representation within a strict latency. Here "long" means the length of sequential user behavior reaches up to 1000 or more.  Practically, the system latency and storage cost increase approximately linear with the length of user behavior sequence. As reported in \cite{zhou2019dien}, it does much engineering work to deploy the sequential model, which just handles user behaviors with maximal length of 50. Figure \ref{fig:ub} shows the average length of user behavior sequences and corresponding CTR model performance in the online display advertising system in Alibaba. Obviously, it is worth to tackle the challenge of long sequential user behavior modeling.

In this paper, we introduce our hands-on practice with the co-design of machine learning algorithm and online serving system. 
We decouple the user behavior modeling module from the whole CTR prediction system and design specific solution correspondingly. 
\textbf{(i) Serving system view}: We design a separate UIC (User Interest Center) module. UIC focuses on online serving issue of user behavior modeling, which maintains the latest interest representation for each user. A key point of UIC is its updating mechanism. The update of user-wise state, depends only on real-time user behavior trigger event, while not the traffic request. That is, UIC is latency free for real-time CTR prediction. 
\textbf{(ii) Machine learning algorithm view}: Decoupling UIC module only cannot handle the storage problem, as it is still quite difficult to store and make inference  with the length of user behavior sequence scaling up to 1000 for hundreds of millions of users. Here we borrow the idea of memory network from NTM \cite{graves2014neural} and propose a novel architecture named MIMN (Multi-channel user Interest Memory Network). MIMN works in an incremental way and can be implemented with UIC module easily. This helps to tackle the storage challenge. Moreover, MIMN improves traditional NTM with two designs of \textsl{memory utilization regularization} and \textsl{memory induction unit}, making it more efficient for modeling user behavior sequences under limited storage and bringing remarkable gain over model performance. 

Theoretically, combining UIC and MIMN provides us with a solution to handle the user interest modeling with unlimited length of sequential behavior data. Our experiments show the superior of proposed solution, on both model performance and system efficiency. To the best of our knowledge, this is one of the first industrial solutions that are capable of handling long sequential user behavior data with length scaling up to thousands.

The main contributions of this work are summarized as follows:
\begin{itemize}
\item We introduce a hands-on practice with the co-design of learning algorithm and serving system for CTR prediction task. This solution has been deployed in a world's leading advertising system and brings us the ability to handle long sequential user behavior modeling.   
\item We design a novel UIC module, which decouples the heavy user interest computation from the whole CTR prediction process. UIC is latency free for traffic request and allows arbitrary complex model calculation which works in an offline mode w.r.t. realtime inference.         
\item We propose a novel MIMN model, which improves original NTM architecture with two designs of \textsl{memory utilization regularization} and \textsl{memory induction unit}, making it more suitable for user interest learning. MIMN is easily to be implemented with UIC server which incrementally updates the user-wise interest representation.          
\item We conduct careful experiments on both public datasets and an industrial dataset collected from the advertising system in Alibaba. We also share our experience of practical issues for deploying the proposed solution in detail.
    We believe this would help to advance the community.    
\end{itemize}

%% file: ch_relatedwork.tex
\section{Related work}
\textbf{Deep CTR Model.} With the rapid development of deep learning, we have made progress in many areas, such as computer vision\cite{huang2016densely}, natural language processing\cite{bengio:attention}. Inspired by these successes, a number of deep-learning-based CTR prediction methods \cite{deep_crossing,youtube:recommend,deep_intent,widedeep} have been proposed. Instead of feature engineering in traditional methodologies, these methods exploit neural network to capture the feature interactions. Although the idea seems to be simple,  these works make a big step forward in the development of CTR prediction task. After that, industry communities pay more attention on model architecture design instead of improving performance by exhausting feature engineering.
Besides learning feature interactions, an increasing number of approaches are proposed to capture user's insight from the rich historical behavior data. DIN\cite{zhou2018deep} points out that user's interests are diverse and vary with items. Attention mechanism is introduced in DIN to capture user's interest. 
DIEN\cite{zhou2019dien} proposes an auxiliary loss to capture latent interest from concrete behavior and refine GRU\cite{chung2014empirical} to model evolution of interest.

\textbf{Long-term User Interest.} \cite{kim2003learning} argues that long-term interest mean general interest, which is back to one's mind and important for personalization. \cite{liu2007framework} proposes to model user's long-term interest on category. \cite{chung2011incremental} incrementally models long-term and short-term user profile score to express user's interest. All of these methods model long-term interest by feature engineering, rather than by adaptive end2end learning. TDSSM\cite{song2016multi} proposes to jointly model long-term and short-term user interests to improve the recommendation quality. Unfortunately, these deep learning based methods, such as TDSSM, DIN, DIEN could hardly be deployed in a real-time prediction server facing extremely long user behavior sequence. The pressure from storage, latency for computation will grow linearly with the length of user behavior sequence. In industrial application, the length of behavior sequence is typically small, e.g. 50, while an active user in Taobao might leave behaviors, such as click, conversion etc., with  length to exceed 1000 within two weeks only.   

\textbf{Memory Network.} Memory network\cite{sukhbaatar2015end, graves2014neural} has been proposed to extract knowledge with an external memory component. This idea has been widely used in NLP, such as question answering system. Several works\cite{chen2018sequential,ebesu2018collaborative,huang2018improving,wang2018neural} utilize memory network for user interest modeling. However, those methodologies neglect long-term interest modeling and practical deployment issue. 

%% file: ch_sysview_pq.tex
\begin{figure*}[t]
 \centering
 \includegraphics[height=2.4in]{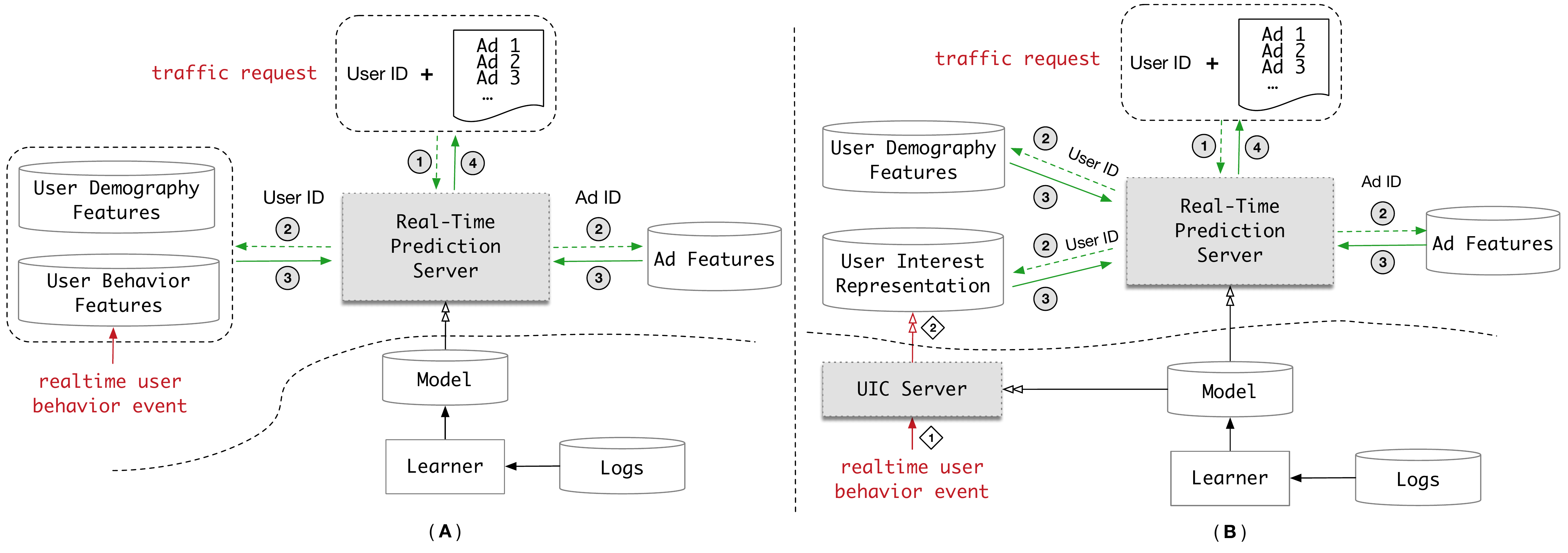} 
 \caption{Illustration of Real-Time Prediction (RTP) system for CTR task. Typically it consists of three key components: feature management module, model management module and prediction server. (A) is the last version of our RTP system and (B) is the updated one with proposed UIC server. The key difference between system A and B is the calculation of user interest representation: (i) In A it is executed within prediction server w.r.t. request. (ii) In B it is executed separately in the UIC server w.r.t. realtime user behavior event. That is, it decoupled away and latency free w.r.t. traffic request. }   
 \label{fig:system}
\end{figure*}

\section{Realtime CTR Prediction System} 
\label{sec:bg}

In real-world recommender or advertising systems, CTR prediction module works as a critical component \cite{youtube:recommend,zhou2018deep}. Usually it receives a set of candidates (e.g. items or ads) and returns correspondingly predicted probability scores by execution of realtime model inference. This process is required to be finished under a strict latency limit, the typical value of which is 10 milliseconds in practice.      

Part A of Fig.\ref{fig:system} gives a brief illustration of \textbf{R}eal\textbf{T}ime \textbf{P}rediction (\textbf{RTP} for short) system for CTR task in our online display advertising system. In order to facilitate the reader to understand, we assume the input of request for RTP includes user and ad information only, omitting context or other factors.

\subsection{Challenges of Serving with Long Sequential User Behavior Data}
\label{sec:challenge}
In industrial applications, such as recommender system in e-commerce industry \cite{zhou2018deep,zhou2019dien}, user behavior features contribute the most volume in the feature set. For example, in our system, nearly $90\%$ of features are user behavior features, with the rest $10\%$ for user demography features and ad features. These behavior data contains rich information and is valuable for user interest modeling~\cite{kim2003learning,liu2007framework,chung2011incremental,song2016multi}. 
Fig.\ref{fig:ub} shows average length of user behavior sequences collected within different days in our system as well as offline performances of basic model (Embedding\&MLP\cite{zhou2018deep}) trained with user behavior features of different length. Without any other effort, basic model with length of 1000 sequence gains $0.6\%$ improvement on AUC compared with the length of 100. It is worth mentioning that $0.3\%$ AUC improvement only is significant enough for our business. The improvement indicates that making use of these long sequential user behavior data is of great value. 

However, exploiting long sequential behavior data brings great challenges. 
 Practically, behavior features from hundreds of millions users are of huge volume. To keep low latency and high throughput in recommender system, behavior features are usually stored in an extra distributed in-memory storage system, such as TAIR\cite{TAIR} in our system. 
These features are fetched to the prediction server and participate in the calculation of realtime inference when receiving traffic request.
  According to our hands-on experience, it costs a lot of engineering work to implement DIEN \cite{zhou2019dien} in our system. Both latency and throughput have reached the performance edge of the RTP system with length of user behavior sequence to be 150, not to mention the case with length of 1000. 
 It is quite difficult to involve more user behavior data directly, as it faces several challenges, of which two most critical ones include:       
\begin{itemize}
	\item \textbf{Storage Constraints.} There are more than 600 millions users in our system. The maximal length of behavior sequences for each user are 150.  It costs about 1 terabyte(TB) storage space which stores not only product\_id but also other related features id such as shop\_id, brand\_id etc. When the length of behavior sequence is up to 1000, 6 TB storage would be consumed, and the number increases linearly with the length of user behavior sequence. As we mentioned before, high performance storage is used in our system to keep low latency and high throughput, and it is too expensive to hold such huge storage. The huge volume also causes corresponding calculation and update of user behavior features to be costly enough. Thus a fairly long behavior sequence means a unacceptable storage consumption.
	\item \textbf{Latency Constraints.} Realtime inference with sequential deep network is well-known to be critically challenging, especially in our scenario with massive request. DIEN \cite{zhou2019dien} is deployed several techniques  to reduce the latency of DIEN serving in our system to 14ms with the QPS (Query Per Second) capacity of each worker to 500.  However, when the length of user behaviors is up to 1000, the latency of DIEN reach to 200 ms with 500 QPS. It is hard to bear in our display advertising system with the latency limitation of 30ms with 500 QPS.
Hence it is impossible to obtain the benefits of long behaviors under present system architecture.
\end{itemize}

\begin{figure*}[t]
 \centering
 \includegraphics[height=3in, width=6.0in]{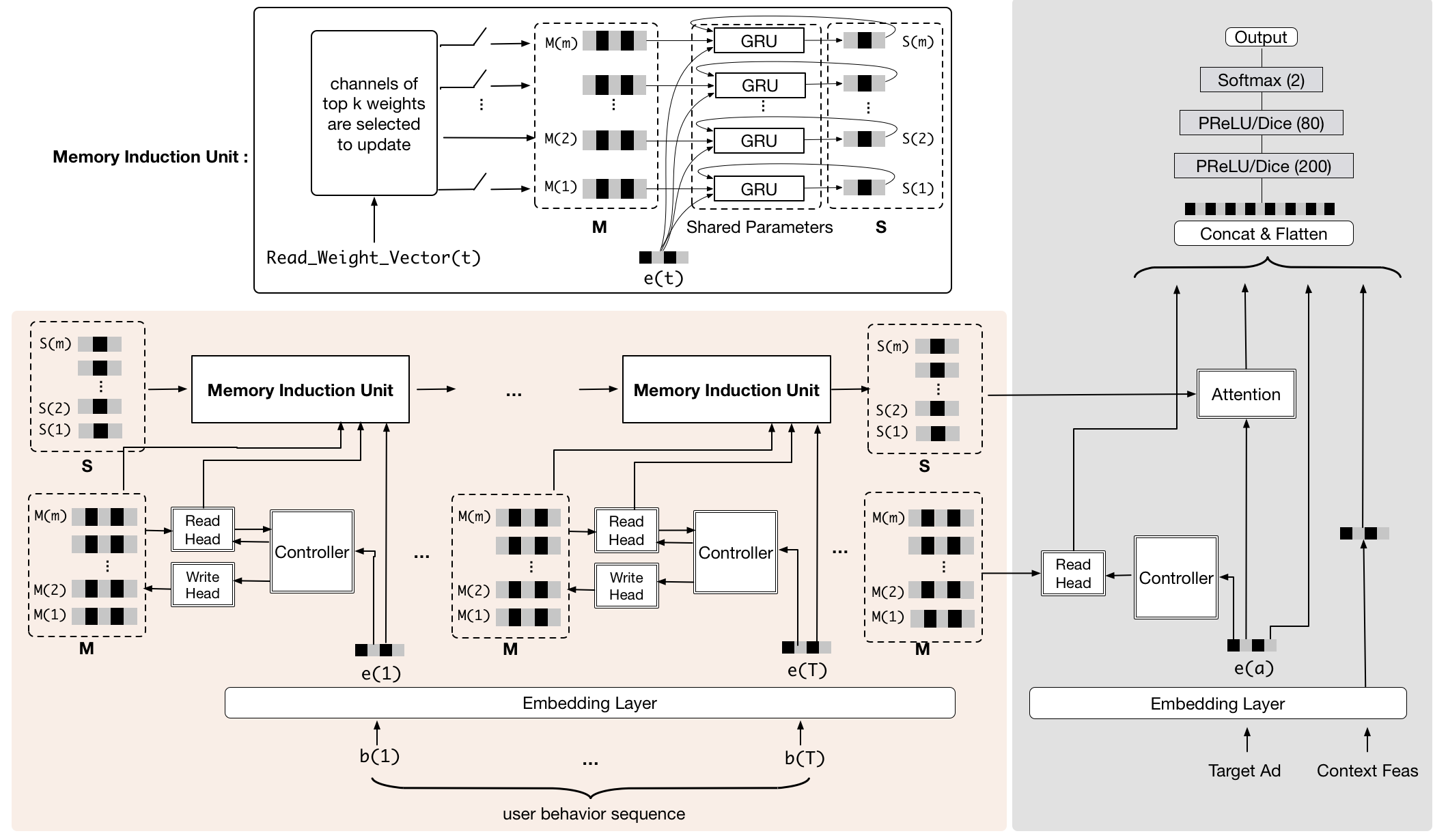} 
 \caption{Network architecture of the proposed MIMN model. MIMN is composed of two main parts: (i) the left sub-network which focuses on user interest modeling with sequential behavior feature, (ii) the right sub-network follows the traditional Embedding\&MLP paradigm which takes the output of left sub-network and other features as inputs. The contribution of MIMN lies in the left sub-network, which is motivated from NTM model and contains two important memory architectures: a) basic NTM memory unit with standard operation of \textsl{memory read} and \textsl{memory write}; b) memory induction unit with multi-channel GRUs to capture high-order information based on the preceding learned NTM memory.  } 
 \label{fig:MIMN}
\end{figure*}

\subsection{User Interest Center}
\label{sec:uic}
To tackle the above mentioned challenges for long sequential user behavior modeling, we propose a solution with the co-design of machine learning algorithm and serving system.  Due to user behavior modeling is the most challenging part of CTR predict system, we design an UIC (User Interest Center) module to handle it.  

Part B of Fig.\ref{fig:system} illustrates the newly designed RTP system with UIC server. The Difference between system A and B is the calculation of user interest representation. In B, UIC server maintains the latest interest representation for each user. A key point of UIC is its updating mechanism. The update of user-wise state, depends only on realtime user behavior trigger event, rather than the request. That is, UIC is latency free for realtime CTR prediction. In our system, UIC can reduce the latency of DIEN model with 1000 length of user behavior from  200 ms to 19ms with 500 QPS.

%% file: ch_approach_gr.tex
\section{Multi-Channel user Interest Memory Network}
In this section, we introduce in detail our machine learning algorithm for long sequential user behavior modeling. 

\subsection{Challenges of Learning From Long Sequential User Behavior Data}
\label{sec:algo_challenge}
Learning from long sequential data is known to be difficult. It is not strange that simple RNNs (RNN\cite{rnn}, GRU\cite{gru}, LSTM\cite{lstm}) fail when facing fairly long sequences. Attention mechanism is introduced to enhance the expressive ability of model by compressing necessary informations of sequential data into a fixed-length tensor\cite{bahdanau2014neural,zhou2018deep}. For example, in DIN\cite{zhou2018deep} attention mechanism works by soft-searching parts of hidden states or source behavior sequences that are relevant to target item. For realtime inference, it needs to store all the raw behavior sequences, which brings great pressure of storage for online system. Besides, the computation cost of attention grows linearly with the length of behavior sequence, which is unacceptable for long sequential user behavior modeling.  
Actually, hidden state in RNNs is not designed to store the whole information of past source sequence but rather to care more about predicting target. Thus, the last hidden state may forget long-term information. Besides, it it redundant to store all hidden states.

Recently, NTM (Neural Turing Machine \cite{graves2014neural}) is proposed to capture information from source sequence and store it in a fixed size of external memory, achieving significant improvement over RNNs models in many tasks of modeling with long sequential data. 

Borrowing the idea from NTM, in this paper we propose a memory network-based model, which provides us a new solution for handling long sequential user behavior modeling. We name this model MIMN (Multi-channel user Interest Memory Network), as illustrated in figure \ref{fig:MIMN}. UIC stores the external memory tensor of MIMN and updates it for each new behavior from user. In this way, UIC incrementally capture user's interests from his/her behavior sequence. Although UIC stores a fixed-length memory tensor instead of raw behavior sequence, the dimension of the memory tensor must be limited when considering the storage pressure,. In this paper, we propose memory utilization regularization to increase the expressive ability of memory tensor in UIC by increasing the utilization of memory. On the other hand, as the user interests vary as well as evolve over time, we propose memory induction unit to help capture high-order information.              

\begin{figure}[t]
 \centering
 \includegraphics[height=1.8in, width=3.2in]{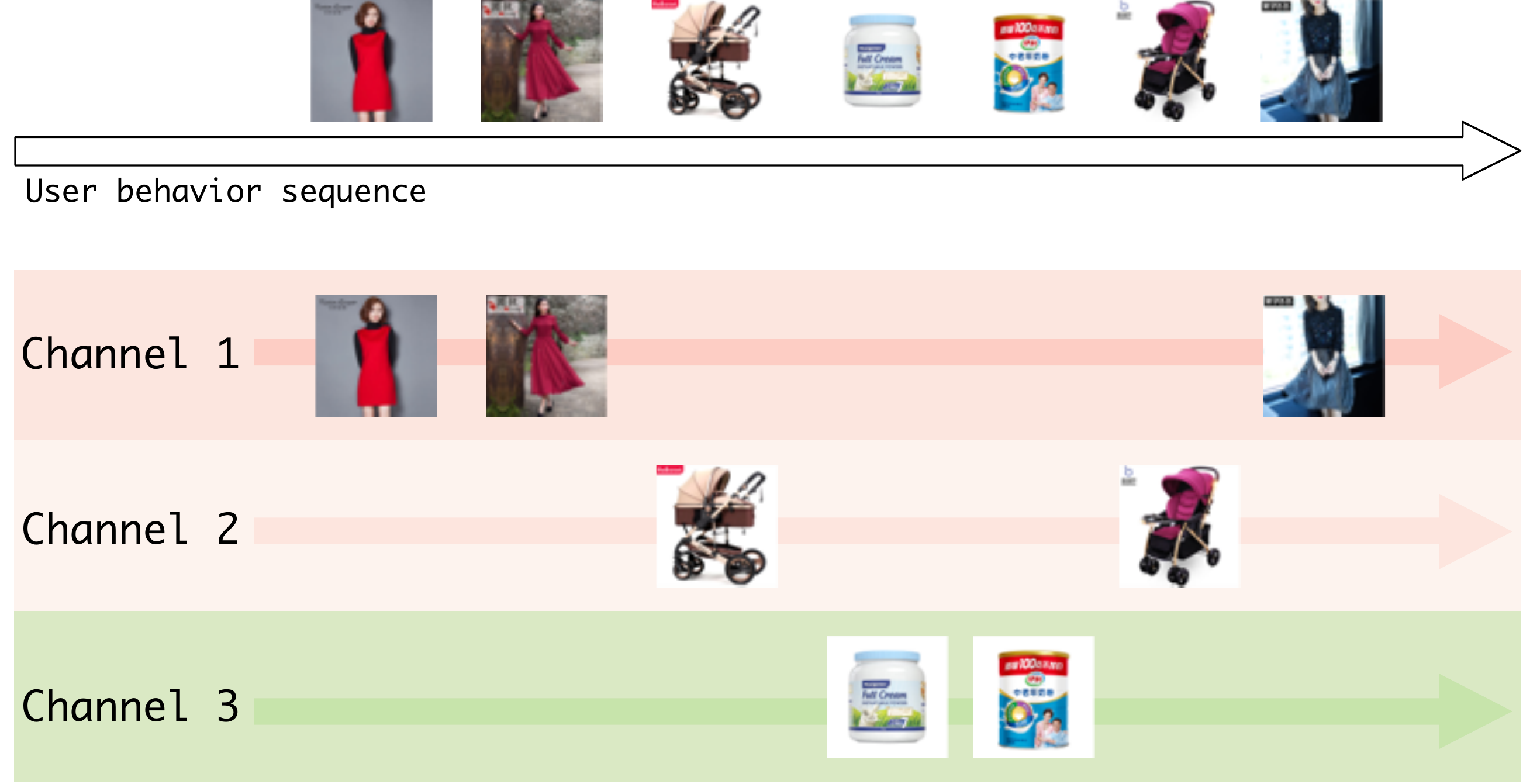}
 \caption{Multi-channel memory induction process. } 
 \label{fig:example} 
\end{figure}

\subsection{Neural Turing Machine}
\label{NTM}
MIMN follows traditional Embedding\&MLP paradigm \cite{youtube:recommend,zhou2018deep,zhou2019dien}, of which for more details we refer the reader to \cite{zhou2018deep}. The structure of MIMN is illustrated in Fig. \ref{fig:MIMN}. 

Standard NTM captures and stores information from sequential data with a memory network. 
In the time step of $t$, parameter of memory is indicated as $\mathbf{M}_t$, which consist of $m$ memory slots $ \{ \mathbf{M}_t(i) \}|_{i=1}^m$.   
Two basic operations for NTM are \textsl{memory read} and \textsl{memory write}, which interact with memory through a controller. 

\textbf{Memory Read.}  Input with $t$-th behavior embedding vector, the controller generates a read key $\mathbf{k}_t$ to address memory. It first traverses all memory slots, generating a weight vector $\mathbf{w}_t^r$ 
\begin{align}
\label{read_weight}
    \mathbf{w}_t^r(i) = \frac{\text{exp}\big(K\big(\mathbf{k}_t, \mathbf{M}_{t}(i)\big)\big)}{\sum_{j}^{m} \text{exp} \big( K\big(\mathbf{k}_t, \mathbf{M}_{t}(j)\big)\big)}, ~~ \textsl{for} ~~ i=1,2,...m 
\end{align}
where \begin{align}    
     K\big(\mathbf{k}_t, \mathbf{M}_{t}(i)\big) = \frac{\mathbf{k}_t^T  \mathbf{M}_{t}(i)}{\parallel\mathbf{k}_t\parallel\parallel\mathbf{M}_{t}(i)\parallel}, 
\end{align}
then calculates a weighted memory summarization as output $\mathbf{r}_t$,
\begin{align}        
    \mathbf{r}_t = \sum_{i}^{m} {w}_t^r(i)\mathbf{M}_{t}(i).
\end{align}
    
\textbf{Memory Write.} The weight vector $\mathbf{w}^{w}_t$ for memory write addressing is generated similar to \textsl{memory read} operation of Eq.(\ref{read_weight}). 
Two additional keys of add vector $\mathbf{a}_t$ and erase vector $\mathbf{e}_t$ are also generated from the controller, which control the update of memory.  
\begin{equation}
\label{eq:soft_write}
\mathbf{M_t} = \mathbf{(1 - \mathbf{E}_t) \odot M_{t - 1}}  + \mathbf{A_t} \; ,
\end{equation}
where $\mathbf{E_t}$ is the erase matrix, $\mathbf{A_t}$ is the add matrix, with $\mathbf{E_t} = \mathbf{w}^{w}_t \otimes \mathbf{e}_t$ and $\mathbf{A_t} = \mathbf{w}^{w}_t \otimes \mathbf{a}_t$. Here $\odot$ and $\otimes$ means dot product and outer product respectively.    

\subsection{Memory Utilization Regularization}
Practically, basic NTM suffers from unbalanced memory utilization problem, especially in the scenario of user interest modeling. That is, hot items tend to appear easily in the sequence of user behavior data and dominate the memory update, making the usage of memory to be inefficient. Previous works \cite{rae2016scaling,santoro2016one} in NLP area have proposed to use the LRU strategy to balance the utilization of each memory. Because LRU pays much attention to balance the utilization of memory in every short period of sequence during the processing, LRU almost never writes information into same slot for adjacent time steps. However, in our scenario, users may interact with a several behaviors belonging to same interest, which therefore should be written into same slot. LRU will disorganize the content addressing and is not suitable for our task. In this paper, we propose a new strategy named memory utilization regularization, which is shown to be effective for user interest modeling.   

\noindent \textbf{Memory Utilization Regularization.} The idea behind memory utilization regularization strategy is to regularize the variance of write weight across different memory slots, pushing the memory utilization to be balanced.
Let $\mathbf{g}_t = \sum_{c=1}^t\mathbf{w}^{\tilde{w}}_c$ be the accumulated update weight till $t$-th time step, where $\mathbf{w}^{\tilde{w}}_c$ represents the re-balanced write weight in c-th time-step. The re-balanced write weight $\mathbf{w}_t^{\tilde{w}}$ can be formulated as:
\begin{align}
&  P_t = {softmax}(W_{g}\mathbf{g}_t) \\
&  \mathbf{w}_t^{\tilde{w}} = \mathbf{w}_t^wP_t.
\end{align}
$\mathbf{w}_t^w$ is the original write weight introduced in subsection \ref{NTM}, and $\mathbf{w}_t^{\tilde{w}}$ represents the new write weight for memory update.
The weight transfer matrix $P_t$ depends on (i) $\mathbf{g}_t$ which represents the accumulated utilization of each memory slot at $t$-th step, (ii) parameter matrix $W_{g}$ which is learned by a regularization loss: 
\begin{align}
    & \mathbf{w}^{\tilde{w}} = \sum_{t=1}^T \mathbf{w}_t^{\tilde{w}},  \\
    & \mathbf{L}_{reg} = \lambda \sum_{i=1}^{m} \big(\mathbf{w}^{\tilde{w}}(i) -  \frac{1}{m}\sum_{i=1}^m{\mathbf{w}^{\tilde{w}}(i)} \big)^2, 
\end{align}
where $m$ is the slot number of memory. $L_{reg}$ helps to reduce the variance of update weight across different memory slots.
Replacing $\mathbf{w}_t^w$ with $\mathbf{w}_t^{\tilde{w}}$, the update rates for all m slots tend be even. In that way, the utilization of all the memory slots are enhanced to be balanced. Utilization regularization can help memory tensor to store more information from source behavior data.

\subsection{Memory Induction Unit} 
Memory in NTM is designed to store original information from source data as much as possible. A fly in the ointment is that it may miss capturing some high-order information, such as the evolving process of each part of interests.
To further enhance the capability of user interest extraction, MIMN designs a Memory Induction Unit (MIU).  
MIU also contains an internal memory $\mathbf{S}$, with the number of slots to be $m$, which is the same as NTM. 
Here we refer to each memory slot as a user interest channel.       
At $t$-th time step, MIU: (i) chooses $k$ channels with channel indexes in set $\{i:\mathbf{w}_t^r(i) \in top_k(\mathbf{w}_t^r)\}|_{i=1}^k$.  $\mathbf{w}_t^r$ is the weight vector for \textsl{memory read} of the aforementioned NTM, shown in Eq.(\ref{read_weight}). (ii) for $i$-th chosen channel, updates $\mathbf{S}_t(i)$ according to Eq.(\ref{eq:miu}). 
\begin{equation}
\label{eq:miu}
\mathbf{S}_t(i) = \text{GRU}(\mathbf{S}_{t-1}(i), \mathbf{M}_t(i),  e_t),
\end{equation}
where $\mathbf{M}_t(i)$ is the $i$-th memory slot of NTM and $e_t$ is the behavior embedding vector.                     
Eq.(\ref{eq:miu}) shows that MIU captures information from both the original behavior input and information memorized in NTM module. This acts as an inductive process, which is illustrated in Fig.\ref{fig:example}. 
Parameters of GRU for multi-channel memory are shared, without increasing the parameter volume.

\subsection{Implementation for Online Serving}
\label{sec:impl}
Different from \cite{zhou2018deep,zhou2019dien} which applies attention mechanism to obtain a candidate-centric interest representation, MIMN learns to capture and store user's diverse interests explicitly in an external memory for each user.
 This memory-based architecture doesn't need interactive computation between candidate (e.g. target ad in our system as illustrated in Fig.\ref{fig:MIMN}) and user behavior sequences, and can be executed incrementally, making it scalable for long sequential user behavior modeling.   

\begin{figure}[t]
 \centering
 \includegraphics[height=2.0in, width=3.2in]{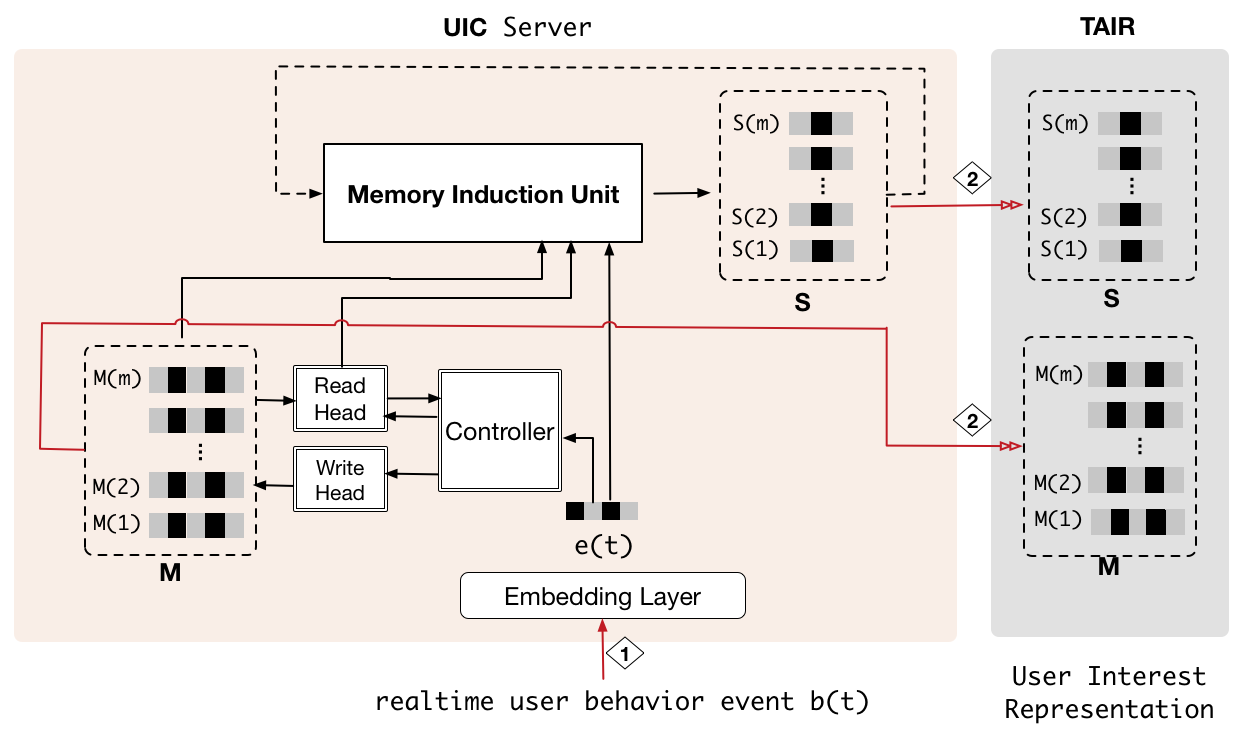}
 \caption{Implementation of the sub-network for user interest modeling with NTM and MIU in UIC Server.} 
 \label{fig:uic}
\end{figure}

The implementation of MIMN for online serving is straight-forward. As was introduced in section \ref{sec:uic}, we split and implement the entire model in two servers: the left sub-network for user interest modeling  of the most heavy computation with NTM and MIU is implemented in UIC server, as illustrated in figure \ref{fig:uic}, leaving the rest of right sub-network to be implemented in RTP server.  Fig.\ref{fig:MIMN} illustrates clearly this implementation.  

Both NTM and MIU module enjoy the benefit of incremental calculation. The latest memory states represent user interests and are updated to TAIR for realtime CTR prediction. When receiving a new user behavior event, UIC calculates and updates the user interest representation to TAIR again.  In this way, user behavior data is not needed to be stored.
 The huge volume of longtime user behaviors could be reduced from 6T to 2.7T in our system.     

\textbf{Discussion.} The co-design of UIC server and MIMN algorithm enables us to handle long sequential user behavior data with length scaling up to thousands. The update of UIC for user-wise interest representation is independent of the whole model calculation, making it latency free for real-time CTR prediction. MIMN proposes to model user interests in an incremental way, without need to store the whole user behavior sequence as traditional solution does. Besides, MIMN is designed with improved memory architectures, enabling superior model performance. However, it is not suitable in all the situations. We recommend to apply this solution in applications with: (i) rich user behavior data, (ii) the scale of traffic volume for real-time user behavior event can not to significantly exceed that of real-time CTR prediction request.

%% file: ch_exp.tex
\section{Experiments}
In this section, experiments are organized in two folds: (i) We present algorithm validation in detail, including the datasets, experimental setup, compared models, and corresponding analysis. Both the public datasets and experimental codes are made available\footnote{https://github.com/UIC-Paper/MIMN}. (ii) We discuss and share our practical experience of deploying the proposed solution in the display advertising system in Alibaba.

\subsection{Datasets and Experimental Setup}
Model comparisons are conducted on two public datasets as well as an industrial dataset collected from online display advertising system of Alibaba. Table \ref{table:Statistics} shows the statistics of all datasets.  

\begin{table}[]
\caption{Statistics of datasets used in this paper.}
\small
\centering
\begin{threeparttable}
\begin{tabular}{lcccc}
\toprule
    Dataset       & Users & Items\tnote{a} & Categories & Instances \\ \midrule
Amazon(Book). & 75053 & 358367 & 1583 & 150016 \\ 
Taobao. & 987994 & 4162024 & 9439 & 987994 \\
Industrial.  & 0.29 billion & 0.6 billion & 100,000 & 12.2 billion \\ 
\bottomrule
\end{tabular}
			\begin{tablenotes}
	        \item[a] For industrial dataset, items refer to be the advertisement.
		    \end{tablenotes}
 \end{threeparttable}
\label{table:Statistics}
\end{table} 

\textbf{Amazon Dataset\footnote{http://jmcauley.ucsd.edu/data/amazon/}} is composed of product reviews and metadata from Amazon \cite{mcauley2015image}. We use the \textsl{Books} subset of Amazon dataset. For this dataset, we regard reviews as one kind of interaction behaviors, and sort the reviews from one user by time. Assuming there are T behaviors of user $u$, our purpose is to use the previews $T-1$ behaviors to predict whether user $u$ will write reviews that shown in $T-th$ review. To focus on long sequence user behavior prediction, we filter these samples whose length of behavior sequence is shorter than 20 and truncate behavior sequence at length 100.

\textbf{Taobao Dataset\footnote{https://tianchi.aliyun.com/dataset/dataDetail?dataId=649\&userId=1}} is a collection of user behaviors from Taobao's recommender system \cite{Han2018Learning}. The dataset contains several types of user behaviors including click, purchase, etc. It contains user behavior sequences of about one million users. We take the click behaviors for each user and sort them according to time in an attempt to construct the behavior sequence. Assuming there are T behaviors of user u, we use the former T-1 clicked products as features to predict whether users will click the $T$-th product. The behavior sequence is truncated at length 200. 

\textbf{Industrial Dataset} is collected from online display advertising system of Alibaba. Samples are constructed from impression logs, with' click' or 'not' as label. Training set is composed of samples from past 49 days and test set from the following day, a classic setting for industrial modeling. In this dataset, user behavior feature in each day's sample contains historical behavior sequences from the preceding $60$ days, with length truncated to 1000.  

\textbf{Experimental Setup}
For all models, we use Adam\cite{kingma2014adam} solver. We apply exponential decay with learning rate starting at 0.001 and decay rate of 0.9. Layers of FCN (fully connected network) are set by $200 \times 80 \times 2$. The number of embedding dimension is set to be 16, which is the same as the dimension of memory slots. The number of hidden dimension for GRU in MIU  is set to be 32. Number of memory slots in both NTM and MIU is a parameter that is  examined carefully in the ablation study section.  We take AUC as the metric for measurement of model performance.

\subsection{Competitors}
We compare MIMN with state-of-the-art CTR prediction models in the scenario of long sequential user behavior modeling. 
\begin{itemize}
        \item{{\it \textbf{Embedding\&MLP}}} is the basic deep learning model for CTR prediction. It takes sum pooling operation to integrate behavior embeddings.
        \item{\it \textbf{DIN}}~\cite{zhou2018deep} is an early work for user behavior modeling which proposes to soft-search user behaviors w.r.t. candidates.         
        \item{{\it \textbf{GRU4Rec}}}~\cite{hidasi2015session} bases on RNN and is the first work using the recurrent cell to model sequential user behaviors.
        \item{{\it \textbf{ARNN}}} is a variation of GRU4Rec which uses attention mechanism to weighted sum over all the hidden states along time for better user sequence representation.
       	\item{{\it \textbf{RUM}}}~\cite{chen2018sequential} uses an external memory to store user's behavior features. It also utilizes soft-writing and attention reading mechanism to interact with the memory. We use the feature-level RUM to store sequence information. 
        \item{{\it \textbf{DIEN}}}~\cite{zhou2019dien}
        integrates GRU with candidate-centric attention trick for capturing the evolution trend of user interests and achieves state-of-the-art performance. For fare comparison, we omit the trick of auxiliary loss for better embedding learning in DIEN, which otherwise should be implemented for all the above mentioned models. 
\end{itemize}

\subsection{Results on Public Datasets}
Table~\ref{tab:public} presents results of all the compared models and MIMN. Each experiment is repeated 3 times. 
\begin{table}
	\small
	\centering
	\caption{Model performance (AUC) on public datasets}\label{tab:public}
	\resizebox{\columnwidth}{!}{%
		\begin{tabular}{l c c}
			\addlinespace
			\toprule
			Model        & Taobao (mean $\pm$ std) & Amazon (mean $\pm$ std)\\
			\midrule
			{\it Embedding\&MLP}~     & $ 0.8709 \pm 0.00184$  &  $ 0.7367 \pm 0.00043 $  \\
			{\it DIN}~     & $ 0.8833 \pm 0.00220$  &  $ 0.7419 \pm 0.00049 $  \\			
			{\it GRU4REC}~ & $ 0.9006 \pm 0.00094$  &  $ 0.7411 \pm 0.00185 $  \\
			{\it ARNN}~    & $ 0.9066 \pm 0.00420$  &  $ 0.7420 \pm 0.00029 $  \\
			{\it RUM}~    & $ 0.9018  \pm 0.00253$  &  $ 0.7428 \pm 0.00041 $  \\
			{\it DIEN}~    & $ 0.9081 \pm 0.00221$  &  $ 0.7481 \pm 0.00102 $  \\
			{\it MIMN}~    & $\bm{0.9179 \pm 0.00325}$  &  $\bm{0.7593 \pm 0.00150}$  \\
			\bottomrule
	\end{tabular}}
\end{table}

All the other models beat Embedding\&MLP, which validates the effectiveness of network architecture design for user behavior modeling. 
MIMN beats all the models with a significant gain over AUC metric. 
We believe this is due to the huge capacity of memory-based architecture that is suitable for user behavior modeling. As is discussed in section \ref{sec:algo_challenge} that user interests behind long sequential behavior data are diverse and evolve over time. MIMN learns to capture user interests with multi-channel memories in two aspects: (i) memory in basic NTM with balanced utilization for interest memorization, (ii) memory in MIU which further captures high-order information by inducing the sequential relationships of interests based on the memory of NTM.

\subsection{Ablation Study}
In this section, we study the effect of different modules in MIMN. 

\textbf{Slot Number of Memory.}
We conduct experiments on MIMN with different number of memory slots, which is is a manual setting. For simplification we evaluate MIMN with the basic NTM architecture only, omitting the design of \textsl{memory utilization regularization} and \textsl{memory induction unit}. Table~\ref{tab:slot} shows the results.      

\begin{table}
	\footnotesize
	\centering
	\caption{Model performance w.r.t. different slot numbers}\label{tab:slot}
	\resizebox{1\columnwidth}{!}{%
		\begin{tabular}{l c c}
			\addlinespace
			\toprule
		~~~~	Model        ~~~~       & Taobao (mean $\pm$ std) &Amazon (mean $\pm$ std) \\
			\midrule
		~~~~	{\it MIMN 4 slot}     ~~~~     & $ 0.9046 \pm 0.00135 $  &  $ \bm{0.7522 \pm 0.00231}$  \\
		~~~~	{\it MIMN 6 slot}      ~~~~    & $ 0.9052 \pm 0.00202 $  &  $ 0.7503 \pm 0.00120$   \\
		~~~~	{\it MIMN 8 slot}      ~~~~    & $ \bm{0.9070 \pm 0.00186} $  &  $ 0.7486 \pm 0.00071$  \\
			\bottomrule
	\end{tabular}}
\end{table}

Empirically, slot number affects the model performance. For Amazon dataset the best performance achieves with 4 slots while for Taobao dataset is 8. Our analysis result indicates that it  is related to the length of user behavior sequence in the datasets. 
Every slot of memory is randomly initialized. For datasets with long behavior sequence, e.g. Taobao dataset, memory has more chances to learn and achieve stable representation. 
In cases with short behavior sequence, e.g. Amazon dataset, model performance with larger memory capacity suffers from the learning. Especially when the utilization of all the slots of memory are unbalanced, part of memory vectors may not be utilized and updated adequately which means these memory vectors are still kept near to the original initialization.
It will hurt the performance of model. Thus we propose Memory Utilization Regularization to alleviate this problem. 

\begin{figure}[!htb]
\includegraphics[width=8.5cm]{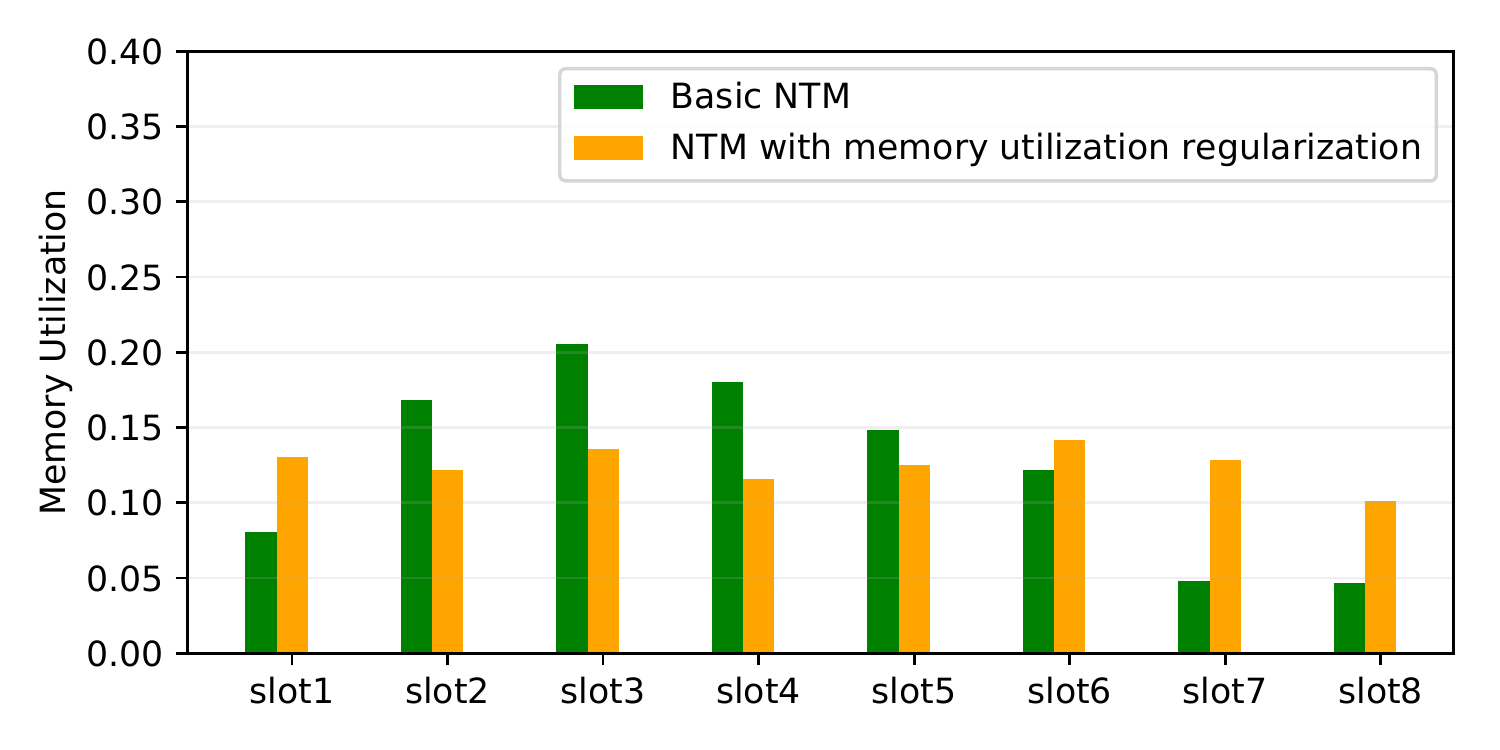}
\caption{Memory utilization over different slots in NTM}
\label{fig:slot_util}
\end{figure}

\textbf{Memory Utilization Regularization.}
With nonuniform interest intensity for each user and the random initialization for memory, the utilization of storage can be unbalanced in basic NTM model. This issue will hurt the learning of memory and make it insufficient for utilizing the limit memory storage. We employ \textsl{memory utilization regularization} trick to help tackle this problem. Figure~\ref{fig:slot_util} shows the memory utilization, which validates the effectiveness of the proposed regularizer. This balanced effect also brings improvement over the model performance, as shown in table~\ref{tab:Multi-Channel}.

\textbf{Memory Induction Unit.} 
\begin{table}
	\small
	\centering
	\caption{Comparison of model performance (AUC) of MIMN with/without memory utilization regularization and memory induction unit}\label{tab:Multi-Channel}
	\begin{threeparttable}
	\resizebox{\columnwidth}{!}{%
		\begin{tabular}{l c c}
			\addlinespace
			\toprule
			Model               & Taobao (mean $\pm$ std) & Amazon (mean $\pm$ std) \\
			\midrule
			{\it MIMN without MUR\tnote{a} ~and~ MIU\tnote{b}}        & $ 0.9070 \pm 0.00186 $  &  $ 0.7486 \pm 0.00071 $  \\
			{\it MIMN with MUR}           & $ 0.9112 \pm 0.00267$  &   $ 0.7551 \pm 0.00121 $  \\
			{\it MIMN with MUR and MIU}     & $ 0.9179 \pm 0.00208$  &   $ 0.7593 \pm 0.00296 $  \\
			\bottomrule
	\end{tabular}}
				\begin{tablenotes}%[para,flushleft]
	        \item[a] MUR stands for \textsl{Memory Utilization Regularization}
	        \item[b] MIU stands for \textsl{Memory Induction Unit}
		    \end{tablenotes}
	\end{threeparttable}
\end{table}
By inducing memory from basic NTM, MIMN with \textsl{memory induction unit} is capable of capturing high-order information and  brings more improvements, as shown in table~\ref{tab:Multi-Channel}. It enhances the capability of user interest extraction and helps modeling user interests from long sequential behavior data. 

\begin{table}
		\tiny
        \centering
        \caption{Model performance (AUC) on industrial dataset}\label{tab:deploy}
        \resizebox{0.95\columnwidth}{!}{%
                \begin{tabular}{l c}
                        \addlinespace
                        \toprule
                     ~~~~~   Model ~~~~~  &AUC      ~~~~~ \\
                        \midrule
                     ~~~~~   {\it DIEN} ~~~~~& $0.6541$  ~~~~~\\
                     ~~~~~   {\it MIMN} ~~~~~& $0.6644$     ~~~~~ \\
                     ~~~~~   {\it MIMN under out-synch setting (within one day) } ~~~~~&$0.6644 $ ~~~~~\\
                    %~~~~~    {\it MIMN with 30 slots of memory}~~~~~ &$0.6663 $~~~~~ \\
                     ~~~~~   {\it MIMN trained with big-sale data} ~~~~~&$0.6627 $~~~~~ \\
                        \bottomrule
        \end{tabular}}
\end{table}

\subsection{Results on Industrial Dataset}
We further conduct experiments on the dataset collected from online display advertisement system of Alibaba. We compared MIMN with the model of DIEN. Table~\ref{tab:deploy} shows the results. MIMN improves DIEN with AUC gain of 0.01, which is significant for our business. 

\begin{figure}[!htb]
	\centering
	\includegraphics[width=8.5cm]{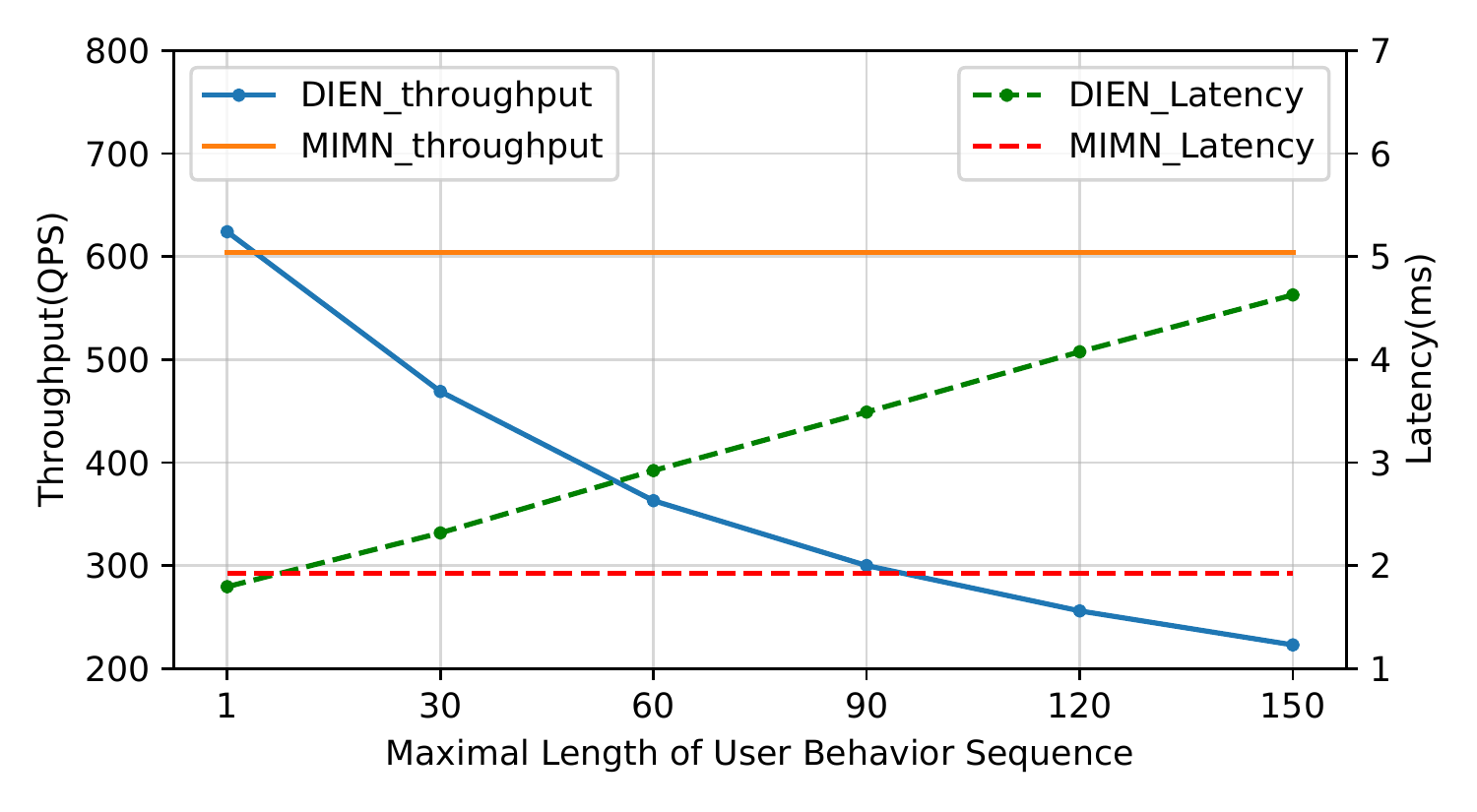}
	\caption{System performance of realtime CTR prediction system w.r.t. different length of user behavior sequences, serving with MIMN and DIEN model. MIMN model is implemented with the design of UIC server.}
	\label{fig:qps}
\end{figure}

Apart from offline model performance, there also exists great difference between MIMN and DIEN model in the aspect of system issue. Figure~\ref{fig:qps} shows the system performance of the real CTR prediction system when serving with models of MIMN and DIEN. MIMN with the co-design of UIC server beats DIEN over a large margin, which holds the property of constant latency and throughput. Hence MIMN is ready to exploit long sequential user behavior data with length scaling up to thousands in our system and enjoys improvement of model performance. On the contrary, system serving with DIEN suffers from both the latency and system throughput. Due to the system pressure, the length of user behavior sequence exploited in DIEN as our last product model is only 50.  This validates again the superiority of our proposed solution.    

\textbf{Online A/B Testing.} 
We have deployed the proposed solution in the display advertising system in Alibaba. 
From 2019-03-30 to 2019-05-10, we conduct strict online A/B testing experiment to validate the proposed MIMN model. 
Compared to DIEN (our last product model), MIMN achieves 7.5\% CTR and 6\% RPM (Revenue Per Mille) gain. We attribute this to the mining of additional information from long sequential behavior data that the proposed co-design solution enables.

\subsection{Practical Experience For Deployment} 
In this section, we discuss practical experience on deploying the proposed solution of UIC and MIMN in our online system.

\textbf{Synchronization of UIC Server and RTP Server}.
As is introduced in section \ref{sec:impl}, MIMN is implemented with UIC and RTP servers together. Hence there is an out-sync problem between UIC and RTP server. Asynchronous parameter update of the two servers might cause incorrect model inference in real systems with periodic model deployment, which is of great risk. Experiment is conducted to simulate the out-sync situation. Table \ref{tab:deploy} shows the results. MIMN with out-sync parameter update shows little difference on model performance. Note in this experiment, the gap of out-sync update time is within one day, which is a traditional setting in industrial systems. Actually in our real system, model deployment is designed to be executed hourly, reducing the risk further. We believe this is due to the stable representation of user interests that MIMN learned, resulting in good generalization performance of MIMN.            

\textbf{The Influence of Big-Sale Data}.
Nowadays, big sale is adopted by many e-commerce sites to attract customers for online consumption, e.g. the famous 11.11 big sale of Alibaba in China. In this extreme situation, the distribution of sample as well as user behaviors are quite different from that in daily situation. We compare the the performance of MIMN with and without training data collected at the big sale day of 11.11 in our system. The results are shown in Table\ref{tab:deploy}. We find it's better to remove the big-sale data empirically. 
    
\textbf{Warm Up Strategy}.
Although UIC is designed to update incrementally, it will take a fairly long time for stable accumulation from the beginning. Practically, we employ the warm up strategy to initialize UIC with pre-calculated user interest representations. That is, we collect historical behaviors of last 120 days (with average length of user behavior sequences to be 1000) for each user and make inference with trained MIMN model in an offline mode, then push the accumulated memories into UIC for further incremental update. This strategy enables a reasonable model performance when deploying the proposed solution as soon as possible.      

\textbf{Rollback Strategy}. 
In case with unexpected problem, e.g. pollution of training samples by large-scale online cheating, the incremental update mechanism of UIC server may suffer a lot. A troublesome challenge is to seek to the point where abnormal case happened. To defend against this risk, we design a rollback strategy that stores the copy of learned user interest representation at 00:00 am every day and record that of the last 7 days.

%% file: ch_concln.tex
\section{Conclusions}
In this paper, we focus on exploiting long sequential user behavior data with the co-design of machine learning algorithm and online serving system. In terms of computation, storage is the main bottleneck for capturing long-term user interests from fairly long sequential user behavior data. We introduce our hands-on practice with the novel solution-a decoupled UIC server for real-time inference of user interest modeling and a memory-based MIMN model that can be implemented incrementally and outperform the other state-of-the-art models. 

It is worth mentioning that deep learning brings us with a powerful toolkit to involve more valuable data for industrial applications. We believe this work opens a new space by modeling with extremely long sequential user behavior data. In the future, we plan to further push forward the research, including learning algorithm, training system as well as online serving system.